\documentclass[particles,article,accept,pdftex,moreauthors]{Definitions/mdpi} 

\firstpage{1} 
\pubvolume{1}
\issuenum{1}
\articlenumber{0}
\pubyear{2025}
\copyrightyear{2025}
\externaleditor{Firstname Lastname} 
\datereceived{30 January 2025} 
\daterevised{21 February 2025} 
\dateaccepted{14 March 2025} 
\datepublished{ } 
\hreflink{https://doi.org/} 

\Title{Muographic Image Upsampling with Machine Learning for Built Infrastructure Applications}

\TitleCitation{Muographic Image Upsampling with Machine Learning for Built Infrastructure Applications}

\Author{William 
 O'Donnell $^{1,2,}$*\orcidA{}, David Mahon $^{1,2}$\orcidB{}, Guangliang Yang $^{1,2}$, Simon Gardner $^{1}$\orcidD{}}

\AuthorNames{William O'Donnell, David Mahon, Guangliang Yang and Simon Gardner}

\AuthorCitation{O'Donnell, 
 W.; Mahon, D.; Yang, G.; Gardner, S.}

\address{%
$^{1}$ \quad School of Physics \& Astronomy, University of Glasgow,
Kelvin Building, University Avenue, \linebreak  Glasgow G12 8QQ, UK; 
 \\
$^{2}$ \quad Lynkeos Technology Ltd, University of Glasgow, No. 11 The Square, Glasgow G12 8QQ, UK}

\corres{Correspondence: w.odonnell.1@research.gla.ac.uk}

\abstract{The civil engineering industry faces a critical need for innovative non-destructive evaluation methods, particularly for ageing critical infrastructure, such as bridges, where current techniques fall short.
Muography, a non-invasive imaging technique, constructs three-dimensional density maps by detecting the interactions of naturally occurring cosmic-ray muons within the scanned volume.
Cosmic-ray muons offer both deep penetration capabilities due to their high momenta and inherent safety due to their natural source.
However, the technology's reliance on this natural source results in a constrained muon flux, leading to prolonged acquisition times, noisy reconstructions, and challenges in image interpretation.
To address these limitations, we developed a two-model deep learning approach.
First, we employed a conditional Wasserstein Generative Adversarial Network with Gradient Penalty (cWGAN-GP) to perform predictive upsampling of undersampled muography images.
Using the Structural Similarity Index Measure (SSIM), 1-day sampled images were able to match the perceptual qualities of a 21-day image, while the Peak Signal-to-Noise Ratio (PSNR) indicated a noise improvement to that of 31 days worth of sampling.
A second cWGAN-GP model, trained for semantic segmentation, was developed to quantitatively assess the upsampling model's impact on each of the features within the concrete samples.
This model was able to achieve segmentation of rebar grids and tendon ducts embedded in the concrete, with respective Dice--Sørensen accuracy coefficients of 0.8174 and 0.8663.
This model also revealed an unexpected capability to mitigate---and in some cases entirely remove---z-plane smearing artifacts caused by the muography's inherent inverse imaging problem.
Both models were trained on a comprehensive dataset generated through Geant4 Monte Carlo simulations designed to reflect realistic civil infrastructure scenarios.
Our results demonstrate significant improvements in both acquisition speed and image quality, marking a substantial step toward making muography more practical for reinforced concrete infrastructure monitoring applications.}


\begin{document}


\section{Introduction}
It has been widely established that the growing quantity of reinforced-concrete-based infrastructure nearing the end of its original intended service life poses a significant challenge.
Current reflection-based near-surface detection techniques---Ground-
Penetrating Radar (GPR) and ultrasonic echo measurements---are limited in scenarios with high concrete thickness and crowded imaging volumes.
While X-ray planar tomography can provide high-resolution imaging at depth, its use is heavily restricted by radiation protection regulations, meaning that it goes largely unused.
Consequently, there are no established Non-Destructive Testing (NDT) technologies capable of inspecting deep into concrete structures to determine the placement of steel rebar reinforcement grids and tendon ducts, as~well as identifying defects, such as honeycombing, tendon duct strand corrosion, and air~voiding.

Muon scattering tomography, an~NDT technique which utilises naturally occurring high-energy (GeV) cosmic-ray muons as its imaging source, has shown promise in addressing this gap.
Successfully deployed in applications such as nuclear waste characterization~\cite{mahon_first---kind_2019}, cargo scanning~\cite{barnes_cosmic-ray_2023}, and~industrial pipe inspection~\cite{martinez_ruiz_del_arbol_applications_2022}, muon scattering techniques have been more recently explored as a tool for civil engineering applications, including the inspection of bridges and other built infrastructure such as historical buildings~\cite{bonomi_cosmic_2019}.
A comparative study by Niederleithinger~et~al.~\cite{niederleithinger_muon_2021} evaluated GPR, ultrasound, X-rays, and muography through the imaging of a 600~kg reference block of reinforced concrete, demonstrating that muography could identify the same features as benchmark X-rays while outperforming the resolution of GPR and ultrasound.
However, challenges remain: muography requires days to weeks to produce an accurate tomograph caused by the low muon flux (1 muon $\mathrm{cm}^{-2}\,\mathrm{min}^{-1}$)  
compared to hours for GPR and ultrasound.
Additionally, detector acceptance and limited vertical resolution caused by the inverse imaging problem also result in noisy images and shadowing artefacts from objects above and below the image~plane.

While computationally complex statistical methods~\cite{schultz_statistical_2007} and pattern recognition techniques, including density-based clustering~\cite{giammanco_cosmic_2024} and support vector machines~\cite{das_muography_2022}, have been proposed to enhance muographic imaging, they often fail to address the core issues of long imaging times, smearing effects, and noisy outputs.
Adjacent fields, such as medical imaging, have overcome some of these issues by using Image-to-Image (I2I) machine learning for image reconstruction and enhancement for Computed Tomography (CT), Magnetic Resonance Imaging (MRI), and Positron Emission Tomography (PET) \cite{armanious_medgan_2020, zhou_review_2021}.
While I2I applications are used to reduce imaging times and doses in hospital imaging, machine learning is also deployed as a detection tool using segmentation to aid medical professionals in the detection of cancers, organs, and polyps~\cite{ma_segment_2024}.

We can apply similar machine learning techniques to our muography images; however, unlike medical imaging, there is a lack of sufficient real-world data, which are required for training models.
To address this, the~Monte Carlo modelling software Geant4 
 \cite{agostinelli_geant4simulation_2003, allison_geant4_2006, allison_recent_2016}, widely used in muography for proof-of-concept and feasibility studies, was employed.
Geant4 allows for the simulation of the detectors, concrete blocks, and muons, bypassing the need for costly experimental setups and their associated lengthy integration times.
Using Geant4 along with the Ecomug cosmic muon event generator~\cite{pagano_ecomug_2021}, a~realistic dataset of high- and low-sampled muographic images of concrete interiors was generated for model training.
Preliminary results demonstrate that a conditional Wasserstein Generative Adversarial Network with Gradient Penalty (cWGAN-GP) model can effectively reduce noise and upsample features such as rebar grids and tendon ducts embedded in concrete.
To analyse the perceptual improvements from upsampling, a~second model was trained using the same architecture for a semantic segmentation task to enable pixelwise object detection, thus providing a quantified breakdown of the enhancements on each concrete feature as a result of the upsampling process.
The models significantly reduced noise, smoothed image edges, and mitigated shadow artefacts caused by low vertical resolution.
This work highlights the potential of machine learning in the post-processing of muographic images, addressing key challenges in the field.
By enhancing the image quality and reducing imaging artifacts, these advancements make muography a more viable and attractive technology for investment and industry~adoption.

\subsection{Muographic Imaging for Built~Infrastructure}
Muon scattering tomography and muon absorption radiography---collectively referred to as muography---are next-generation imaging techniques that leverage naturally occurring, high-energy cosmic-ray muons.
These offer a novel NDT solution that overcomes the poor depth resolution and safety concerns of existing methods.
Absorption measurements utilise the attenuation of cosmic-ray muons as they pass through matter, with~lower detection rates indicating higher density regions of the imaged volume.
This technique only requires a single tracker (consisting of a minimum of two detection planes) adjacent to or behind the target volume.
Therefore, it is suited to a wide variety of tasks, having been demonstrated on built infrastructure---such as railway tunnels~\cite{thompson_muon_2020}, dams~\cite{olah_structural_2023}, and reactor buildings~\cite{fujii_investigation_2020}---as well as larger structures such as volcanoes~\cite{tanaka_radiographic_2014} and pyramids~\cite{morishima_discovery_2017}.

Scattering tomography (hereafter referred to as muography) examines the multiple Coulomb scattering of muons as they interact with atomic nuclei while traversing matter.
This technique requires a minimum of four detector planes arranged as two pairs on either side of the volume of interest, which limits its practical applications, yet it remains preferred over absorption methods due to its superior spatial resolution and detail.
By analyzing the variance in the muon scattering angle, the~radiation length of the material along the muon's path can be estimated.
The true distribution of muon multiple scattering angles follows a Cauchy--Lorentz profile, but~a Gaussian approximation can effectively describe 98\% of the distribution, leading to the Rossi formula~\cite{rossi_cosmic-ray_1941} for the standard deviation of muon multiple scattering angles:
\begin{linenomath}
\begin{equation}\label{rossi}
	\sigma_{\theta} \approx \frac{15\,\mathrm{MeV}}{\beta c p} \sqrt{\frac{L}{X_{0}}},
	\end{equation}
\end{linenomath}
where $\beta c$ is the velocity ($\beta = 1$), $L$ is the length of matter traversed by the muon, and $X_{0}$ 
 is the radiation length of the material.
The measurement of the momentum $p$ (in MeV/c) of a detected muon is dependent on the muography detector, with~cost-effective detectors electing to use the average muon momentum (3--4$\,\mathrm{GeV/c}$) over measurement.
Simplifying Equation~(\ref{rossi}) for a non-momentum measurement, we find that the variance of the muon scattering angles is inversely proportional to the radiation length along the muons path:
\begin{linenomath}
\begin{equation}\label{prop}
	\sigma_{\theta}^{2} \propto \frac{L}{X_{0}}.
	\end{equation}
\end{linenomath}
Positional 
detector hits can be used to calculate ingoing and outgoing muon vectors as they traverse the VoI, allowing the scattering angle to be determined.
Using the above relationships, material composition can then be inferred from the scattering~angles.

Muography is inherently an inverse imaging problem, as~due to multiple scattering, it is impossible to infer the exact path taken by the muon through the VoI.
Various reconstruction algorithms have been developed over the years---detailed reviews of which can be found at~\cite{yang_novel_2019, ughade_performance_2023}---however, the Point of Closest Approach (PoCA) algorithm~\cite{schultz_cosmic_2003} remains the most widely used for its simplicity and computational efficiency.
Like most reconstruction algorithms, the PoCA assumes one scattering event per muon, which is calculated as the angle between the incoming and outgoing muon vectors, with~the scattering point placed halfway along the vector describing the distance of closest approach between these vectors.
Due to low muon flux and the ignorance of multiple scattering by these algorithms, accurate tomographs require many muon events.
Therefore, integration times, of~the order of days and weeks, are required in order to gather enough muons events to statistically and sufficiently resolve objects embedded within thick structures such as~concrete.

\subsection{Convolutional Machine Learning Techniques for Image~Processing}
Supervised machine learning has emerged as a powerful data-driven tool capable of constructing models that are trained to generate outputs $\hat{y}$, which approximate ground truths defined as $y = f(x)$.
Machine learning is often only applied when the problem is too complex for conventional methods to solve, where it is able to identify obscure patterns and structures inherent to the~dataset.

Model training consists of two key steps: the forward pass and the backward pass.
During the forward pass, inputs $x$ are passed through the parameters of the model to produce predictions $\hat{y}$.
The accuracy of these predictions---measured by their similarity to the ground truth $y$---is quantified by a differentiable loss function $\mathcal{L}(y,\hat{y})$, which evaluates the error between the predicted and actual outputs.
This is followed by the backward pass, where the gradient of the loss function with respect to each model parameter is calculated.
These gradients are then used by an optimization function to update model parameters accordingly.
This process is repeated thousands of times during training, with~model parameters being iteratively adjusted to minimise the global loss to~achieve $\hat{y} \approx y$.

A trained model can, at~best, only generalise to the dataset on which it was trained; therefore, it is crucial that the dataset accurately reflects the inputs that the fully trained model is expected to encounter.
A well-trained model relies on a dataset that is representative of unseen data; in addition to being diverse, expansive, and sufficiently large to prevent underfitting, the~model fails to capture necessary features, resulting in poor performance on new data.
Conversely, overfitting is often caused by an overly complex model which learns the specifics of the training dataset well, thus failing to generalise to unseen data.
However, we can penalise overfit models so that they are more generalised---called regularisation---using common techniques such as non-linear activation functions, batch normalization, and dropout.
Model performance is assessed on an unseen dataset (validation dataset) for continued assessment of the model generalization over the course of training, which is used to assess and guide hyperparameter tuning.
Once training is complete, the~fully trained model is assessed on the test dataset---independent of the training and validation sets---to compare the performance of different models and~architectures.

There are two primary mechanisms that can be used within image processing: convolution and attention.
Convolutions have long been a cornerstone of image processing even before the rise of modern machine learning, enabling the extraction of localised features from image data.
These methods rely on predefined kernels to generate feature maps from input images. For~instance, Sobel filters~\cite{sobel_isotropic_2014} can be commonly used to detect vertical and horizontal edges in images.
Convolutional Neural Networks (CNNs) build upon this concept by using learnable kernels consisting of weight parameters which are tuned during training to extract intricate features and patterns from the data.
CNNs typically consist of cascading convolutional layers, each containing multiple filters which act on the feature maps produced by the previous layers.
This iterative process enables CNNs to learn increasingly abstract and detailed features, making them highly effective for a wide range of image processing tasks, including denoising, segmentation, classification, object detection, super-resolution, data fusion, and image synthesis.
However, CNNs have a fundamental limitation: their small receptive field.
The information accessible to any one component is restricted to the spatial extent of the square convolutional kernel, which is typically sized between 3~$\times$~3 and 7~$\times$~7 pixels.
Consequently, CNNs can only process localised areas of the input image at a time.
In contrast, attention mechanisms can capture a global context, enabling the model to account for long-range dependencies across an image.
Despite their advantages, attention mechanisms are more complex than CNNs and, given their relative recency, are less tested and explored in the imaging domain.
For this initial study on the application of machine learning to muography, the~focus will remain on well-established CNN architectures due to their proven reliability and effectiveness in the image processing~domain.

\subsubsection*{The 
 Conditional Generative Adversarial Network (cGAN)}

An encoder--decoder network is a model methodology that is widely used for CNNs.
The encoder aims to learn the features that should be extracted from the input image, where the decoder aims to build the feature maps back up to an output image, enhancing relevant features and reducing redundant information.
However, this is a lossy process so information deemed redundant to the network is lost, resulting in a loss of finer~detail.

In order to address the loss of detail in encoder--decoder convolutional networks, a~2015 paper by Ronneberger~et~al.~\cite{navab_u-net_2015} proposed a new architecture called a U-Net.
They solved the problem of information loss by introducing skip connections between the encoder and decoder layers by cropping feature maps from the encoder and concatenating them to complimentary-sized feature maps in the decoder.
This allows for minor details to be preserved, producing a higher quality output image.
This architecture is widely used in image-to-image translation tasks~\cite{siddique_u-net_2021} due to its simplicity when compared to alternative~designs.

A conditional Generative Adversarial Network (cGAN) consists of a generator that produces outputs and a discriminator that moderates these outputs during training, resulting in more accurate results than using a standalone U-Net generator.
Whereas cGANs are conditional on an input, GANs by contrast are unsupervised generative models that aim to create data from a source of random noise.
A standard GAN comprises a generator, which transforms the noise into an output, and~a discriminator, which evaluates whether the output has generated data or part of the real dataset.
The generator’s goal is to produce data that are indistinguishable from the real data, while the discriminator strives to correctly identify whether the data are real or fake.
This dynamic creates an adversarial process, with~the generator attempting to `fool' the discriminator and the discriminator learning to better distinguish real from fake data. 

The back-and-forth continues during training until a Nash equilibrium is reached---a state in game theory where neither player (the generator or the discriminator) can improve their position without affecting the other.
In this context,~Nash equilibrium is achieved when the generator produces outputs that are realistic enough that the discriminator can not reliably differentiate between real and fake data and is forced to guess.
Despite their powerful capabilities, GANs are notoriously difficult to train, often suffering from issues such as mode collapse (where the generator produces limited variability) or imbalance between the generator and discriminator (where one dominates the training process).
To mitigate these challenges, techniques like the Wasserstein GAN and gradient penalty are frequently employed, improving training stability and~convergence.

\section{Materials and~Methods}
Training a supervised machine learning model to upsample muography images of concrete blocks requires a large dataset of paired images.
Therefore, we must first create a dataset that contains different sizes, orientations, and types of rebar grids and tendon ducts, as~well as a variety of different equivalent sampling times.
Spherical air voids were also added to the geometry design in order to assess detection ability---as a precursor to a complimentary study to look at defects---due to their simplicity and small~size.

The design of the Geant4 simulations to produce a large and varied dataset of 2D muography images from 700 unique concrete block designs is outlined below in Section~\ref{simulation_design_section}.
Each concrete block was simulated using 100 days worth of muons, resulting in a total dataset size of 70,000~500~$\times$~500 images, each with 100 different versions corresponding to equivalent sampling times of 1 day to 100 days.
The total image dataset therefore contained seven million images, requiring over a trillion muon events to be simulated.
Simulating a maximum equivalent sampling time of 100 days was arbitrary but intentionally overestimated for this initial study to ensure sufficient sampling for a clear ground truth; by comparison, the~concrete block in Niederleithinger~et~al.~\cite{niederleithinger_muon_2021} was imaged for only 50 days.
These data were used to train the cWGAN-GP models (outlined below in Section~\ref{wgan_gp_section}) using the 100-day image as the ground truth and the input being randomly selected from one of the equivalent sampling times of 1--99 days, ensuring image sampling time~generalization.

One challenging task was to assess the perceptual quality of the output images.
While there are a wide variety of metrics that can be used to assess global image quality of outputs, such as the Structural Similarity Index Measure (SSIM) and Peak Signal-to-Noise Ratio (PSNR), these metrics do not give a detailed breakdown about how the model is affecting the perception of features, such as rebar grids and tendon ducts.
If we want to know how the upsampling model is affecting the perception of features within the concrete, we must find a way of separating features from one another.
A solution to this problem is to train a second model to perform semantic segmentation, which is a~machine learning technique used to classify pixels within an image and assign them to a specific class.
This means that each pixel in the muographic image is assigned to either concrete, rebar, tendon duct, air void, or `unknown'.
`Unknown' objects were added as the fourth and final object type such that the segmentation model could assign a class to miscellaneous objects it deemed to not fall into any of the other classes.
The shape, size, and materials used for `unknown' objects are less constrained; hence, a variety of shapes and densities were used to define them.
To train a model to produce segmentation maps, a~set of images detailing `true' object placement of features is required.
Due to the use of Geant4 Monte Carlo simulations, `true' geometry information is available to generate the segmentation ground truths.
For simplicity reasons, all components of the tendon ducts were assigned to one~class.

Assessing the performance of segmentation models is much less ambiguous than that of an I2I task like denoising or upsampling, as~each class can be split up into multiple binary masks from which to calculate metrics.
One of the most popular metrics for the assessment of segmentation models is the Dice--Sørensen coefficient~\cite{dice_measures_1945, sorensen_method_1948} (hereafter referred to as Dice), which is used to gauge the similarity between two sets of data:
\begin{linenomath}	\begin{equation}\label{dice_coefficient_one_class}
		\mathrm{Dice} = \frac{2 TP}{2TP + FP + FN},
	\end{equation}
\end{linenomath}
where $TP$ defines the number of true positive pixels, $FP$ defines the number of false positive pixels, and~$FN$ defines the number of false negative pixels in a given binary~image.

\subsection{Simulating a~Dataset} \label{simulation_design_section}
The detector system used in the simulations was that of the Muon Imaging System (MIS) in place at the University of Glasgow, which is a scintillating-fibre tracking system~\cite{clarkson_design_2014}.
It consists of four detection modules, each consisting of two orthogonal planes of 1024 \linebreak  2~millimetre-pitch polystyrene-based plastic scintillating fibres with polymethylmethacrylate optical cladding coupled to Hamamatsu H12700A multi-anode Photomultiplier Tubes (PMTs).
When a muon passes through a fibre, it imparts energy, causing photoluminescence which is detected by the PMTs.
A muon track was recorded when coincident muon hits were detected in all four detector planes.
The MIS has an imaging area of 1066~mm~$\times$~1066~mm and was set up with a vertical spacing of 530~mm between the upstream and downstream modules, between~which the concrete block samples of 1000~mm~$\times$~1000~mm~$\times$~200~mm were placed.
Within this block, the~four different object types were placed: rebar grids, tendon ducts, air voids, and `unknowns'.
The subsections immediately below outline the parameters which constrained individual object design and placement by a concrete interior randomisation algorithm.
Objects' parameters were first generated---through sampling from uniform distributions---and then placed in order of rebar--duct--void--unknown, where objects would only be placed if they fit within the boundaries of the concrete block and did not overlap other objects.
In order to preserve the uniform distribution of the number of objects per volume, overlapping objects had their parameters re-sampled until they were successfully placed or the number of placement attempts exceeded~1000.

This setup was modelled in Geant4, where Ecomug was used as the cosmic-ray event generator.
A generating plane was placed directly above the detector from which only muons were generated at randomised positions.
The event generator was configured with its default settings, where the angular distribution was defined with a zenith angle $\theta$ spanning from $0$ to $\pi/2$ and an azimuthal angle $\phi$ ranging from $0$ to $2\pi$. 
The charge ratio of positive to negative muons was set to 1.28, which followed a momentum distribution ranging from 10 MeV to 1~TeV.

A total of 700 concrete samples were each imaged for 100 days worth of muons, with~resultant vector pairs processed using the PoCA to calculate scattering angles.
The resultant volume of scattering angles was subsequently voxelised into 2~mm cubes.
Image slices of the X--Y plane for each voxelised geometry were saved for intervals of one day of sampling.
This resulted in the creation of 70,000 unique images, of~which there were \linebreak  100 different versions to reflect each cumulative day of~sampling.

\subsubsection{Carbon-Steel Rebar~Grids}
The cylindrical steel rods that make up rebar grids were defined in Geant4 to have a uniform density of 7.84~$\mathrm{g}\cdot\mathrm{cm}^{-3}$, with~typical chemical composition within the limits of British standard grade 500B BS 4449.
The diameters of these rods were kept the same for all rods within the same grid, which were chosen from standard rebar sizing diameters of 8, 10, 12, 16, 20, and 25~mm.
Square grid spacing (100~mm, 150~mm, 200~mm, 250~mm) between rods was also randomised, of~which there were a random number of between 2 and 12 in the X or Y direction.
Between one and four different rebar grids were positioned parallel to the X--Y plane for each concrete block configuration.
The minimum of a single grid here ensured that each configuration had at least one~feature.

\subsubsection{Tendon~Ducts}
Tendon ducts consisted of large cylindrical objects that were placed such that they spanned the 1000~mm volume in either the X--Z or Y--Z planes.
They contained three subcomponents: casing, grout, and steel strands.
The casing material was randomly chosen between a 0.5~mm thick galvanised carbon-steel casing (7.97~$\mathrm{g}\cdot\mathrm{cm}^{-3}$), 3~mm thick High Density Polyethylene (HDPE) with a density of 0.94~$\mathrm{g}\cdot\mathrm{cm}^{-3}$, or 3~mm thick High Density Polypropylene (HDPP) with a density of 0.90~$\mathrm{g}\cdot\mathrm{cm}^{-3}$.
Casing diameter was sampled from from that of 50, 60, 70, 80, 90, and 100~mm, a~value which also determined the number of high tensile steel strands (density of 7.85~$\mathrm{g}\cdot\mathrm{cm}^{-3}$) contained within the duct.
The grout which filled the space inside the casing had a bulk density of 1.30~$\mathrm{g}\cdot\mathrm{cm}^{-3}$ using British Standard Grade C40/50 BS~8500-1.

\subsubsection{Air~Voids}
Air pockets caused by improper concrete pouring are a defect that can reduce the structural integrity of reinforced concrete~\cite{sun_air_2022}.
Due to the simplicity of this defect, spherical air voids were placed in the concrete in order to gauge muography's resolving capacity in~lieu of a future study.
Each volume contained 0--3 spherical air voids, with~diameters ranging from 10 to 100~mm.

\subsubsection{`Unknown'~Objects}
Between 0 and 2 `unknowns' were added as a control to the dataset consisting of objects we would not find in reinforced concrete to add variety and shape for model generalization.
These consisted of boxes, cylinders, and spheres which did not exceed a 3D bounding box with edges limited between 35~mm and 75~mm.
The densities of these objects were randomly selected from water, aluminium, iron, lead, and uranium and were randomly orientated upon placement in the~volume.

\subsection{Pix2pix with~WGAN-GP} \label{wgan_gp_section}
The pix2pix model is a conditional Generative Adversarial Network (cGAN) that utilises a U-Net for its generator and a PatchGAN for its discriminator~\cite{isola_image--image_2017}. In~the base pix2pix cGAN model, the~generator minimises a combined loss:
\begin{linenomath}
\begin{equation}\label{pix2pix_g_loss}
		\mathcal{L}_{\text{G}} = \mathcal{L}_{\text{adv}} + \lambda_{\text{pixel}} \cdot  \mathcal{L}_{\text{pixel}} \text{,}
	\end{equation}
\end{linenomath}
where \( \mathcal{L}_{\text{pixel}} \) is the pixelwise loss function of the generator, $\lambda_{\text{pixel}}$ is the weighting of the pixelwise loss, and \( \mathcal{L}_{\text{adv}} \) is the adversarial loss based on binary classification in the discriminator (real/fake).
This binary classification provides the generator with little information on how close the distribution of generated images is to true distribution.
To better quantify this difference, we can use the Wasserstein distance, also known as the Earth mover's distance~\cite{arjovsky_wasserstein_2017}.
This measures the performance of the discriminator, increasing the smoothness of gradients and importantly providing informative feedback for training the generator.
Unlike binary classification, the~Wasserstein distance is computed using the Kantorovich--Rubinstein duality, which restricts the discriminator to be a 1-Lipschitz function. This constraint ensures that the discriminator's gradients are stable and meaningful, which is important for the generator to learn~effectively.

The gradient penalty is the preferred solution for satisfying the Lipschitz constraint in WGANs~\cite{gulrajani_improved_2017}.
It directly penalises the gradients of the discriminator to ensure they remain close to 1.
Specifically, the~gradient penalty is computed as follows:
\begin{linenomath}
\begin{equation}\label{gp_loss}
		\mathcal{L}_{\text{GP}} =  \mathbb{E}_{\hat{x}} \left[ (\|\nabla_{\hat{x}} D(\hat{x})\|_2 - 1)^2 \right] \text{,}
	\end{equation}
\end{linenomath}
where \( \hat{x} \) represents interpolated samples between real and generated data, and \( D(\hat{x}) \) is the output of the discriminator.
By applying this penalty, the~discriminator is constrained to satisfy the 1-Lipschitz condition, ensuring stable gradients and improving the overall training stability.
This leads to smoother, more informative gradients for the generator, enabling it to produce higher quality images.
In the case of WGAN-GP, the~generator's loss function, $\mathcal{L}_{\text{G}}$, is updated to include the gradient penalty:
\begin{linenomath}
\begin{equation}\label{wgan_gp_loss}
		\mathcal{L}_{\text{G}} = \mathcal{L}_{\text{adv}} + \lambda_{\text{pixel}} \cdot \mathcal{L}_{\text{pixel}} + \lambda_{\text{GP}} \cdot \mathcal{L}_{\text{GP}} \text{,}
	\end{equation}
\end{linenomath}
where \( \mathcal{L}_{\text{adv}} \) is the adversarial loss based on the Wasserstein distance, \( \mathcal{L}_{\text{pixel}} \) is the pixelwise loss, and \( \mathcal{L}_{\text{GP}} \) is the gradient penalty term which is weighted by $\lambda_{\text{GP}}$.

The U-Net generator architecture remains similar to pix2pix, with~five encoder and decoder blocks each consisting of 4~$\times$~4 convolution---batch normalization---ReLU activation.
The upsampling model's generator pixelwise loss uses the mean absolute error (MAE or L1) loss:
\begin{linenomath}
\begin{equation}\label{l1_loss}
	\mathcal{L}_{\text{L1}} = \frac{1}{N} \sum_{i=1}^N \| \hat{y}_i - y_i \| \text{,}
	\end{equation}
\end{linenomath}
where $N$ is the number of samples, and $\hat{y}_{i}$ and $y_{i}$ are the predicted value and ground truth for the {i{th}} sample.
The segmentation model uses a custom pixelwise loss function of an evenly weighted cross-entropy loss and Dice loss (which is one minus the Dice score),
\begin{linenomath}
\begin{equation}\label{custom_loss}
		\mathcal{L}_{\text{custom}} = \frac{1}{2} \mathcal{L}_{\text{CE}} + \frac{1}{2} \mathcal{L}_{\text{Dice}} \text{,}
	\end{equation}
\end{linenomath}
which balances pixelwise prediction accuracy (via cross-entropy) and structural similarity (via Dice).
Since the background class (concrete) makes up a significant proportion of all pixels, the~model is biased to focus on these pixels.
Therefore, the~Dice loss calculation excludes this class, consisting of evenly weighted contributions of the four object classes: rebar grids, tendon ducts, air voids, and `unknowns'.
Both models were trained for \linebreak  100 epochs (one epoch is a full iteration of the training set), with~a batch size of 48.
The upsampling model training set had each of its input images re-sampled from one of its \linebreak  100 versions at the start of each epoch in~order to achieve generalization to different sampling times.
The upsampling model generates outputs $\hat{y}$ that approximate the expected outputs $y$ of 100 days of sampling.
We therefore trained the segmentation model to take inputs $y$ instead of upsampled images $\hat{y}$, as~we aimed to assess the performance of the $\hat{y}$ outputs without introducing any bias from the upsampling process.
An initial learning rate of 0.01 was used for the generator and 0.001 for the discriminator, after~which a step scheduler was used to steady model convergence, thus reducing the learning rates by an order of magnitude every 25~epochs.


\section{Results}
\unskip

\subsection{Upsampling Model~Results}
The upsampling model aims not only to enhance the perceptual characteristics of low-quality muographic images but also to reduce image noise---a key issue identified in~\cite{niederleithinger_muon_2021} when comparing muography to other NDT techniques.
To evaluate the quality difference between images, we used two common computer vision metrics: the Structural Similarity Index Measure (SSIM) \cite{wang_image_2004}, which assesses perception, and~the Peak Signal-to-Noise Ratio (PSNR), which is widely used in signal processing and imaging to quantify image noise.
These metrics provide insights not only on the upsampling model's performance but also on the quality of the input muography images as they are sampled over~time.

\begin{figure}[H]
	\includegraphics[width=\textwidth]{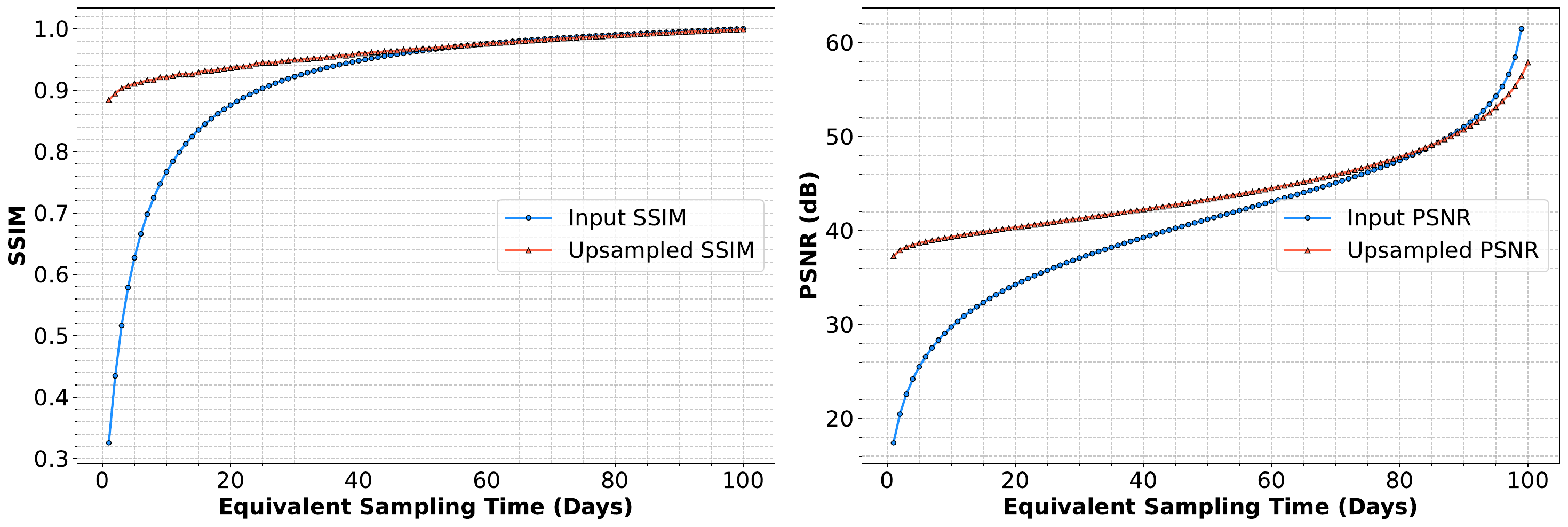}
	\caption{Average Structural Similarity Index Measure (SSIM) and Peak Signal-to-Noise Ratio (PSNR) metrics of the test dataset as~a function of equivalent sampling time. The~original dataset performance is indicated by blue circular points, where the upsampled outputs of the model are indicated by red triangular~points.\label{ssim_psnr}}
\end{figure}

\begin{figure}[H]
	\includegraphics[width=0.9\textwidth]{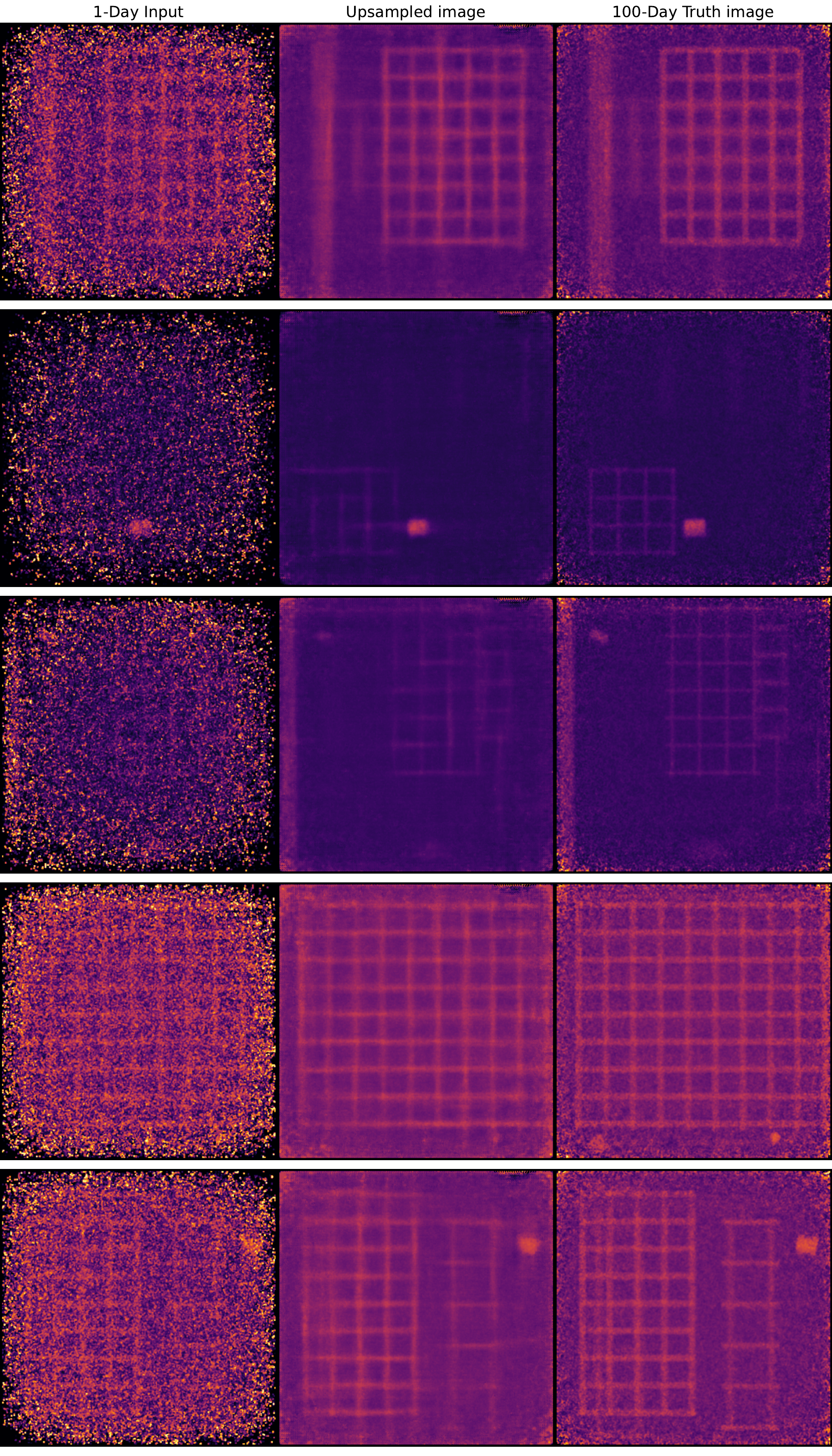}
	\caption{Five 
 examples of 1-day input images from the test dataset (\textbf{left} column) and their corresponding output from the upsampling model (\textbf{middle} column). The~100-day ground truth image for each example is shown for comparison (\textbf{right} column).\label{upsampling_ims}}
\end{figure}

Utilizing the different equivalent sampling time versions for each image in our dataset, we used the aforementioned metrics to look at (a) the effect of the muography sampling time with respect to the image quality and (b) the performance of the upsampling model on different sampling times.
We averaged the SSIM and PSNR metrics of the 6900 images in the test dataset---calculated with respect to the 100-day ground truth---for both the inputs and outputs of the model at each sampling time.
The results are shown in Figure~\ref{ssim_psnr}.
We see that the 1-day upsampled images exhibited SSIM and PSNR scores of 0.88 and 37dB, respectively.
This is equivalent to the perceptual quality of a 21-day unaltered image and the noise quality of a 31-day unaltered image.
However, improvements diminished as the sampling time increased, with~both the input and upsampled images for the SSIM and PSNR converging at an equivalent sampling time of 55 and 85 days, respectively.
Visual inspection of the images before and after upsampling confirmed the improvements indicated by the SSIM and PSNR metrics, even at the lowest quality images---from the 1-day equivalent sampling~time.
Demonstrating these drastic improvements, Figure~\ref{upsampling_ims} compares five 1-day undersampled X--Y plane images to their respective outputs from the upsampled model, as~well as their associated 100-day ground truth values for reference. 

These results confirm that the upsampling model does an excellent job of identifying feature patterns across all sampling times to create clearer and more refined representations of undersampled muography~images.

\subsection{Segmentation Model~Results}
\begin{figure}[H]
	\includegraphics[width=0.9\textwidth]{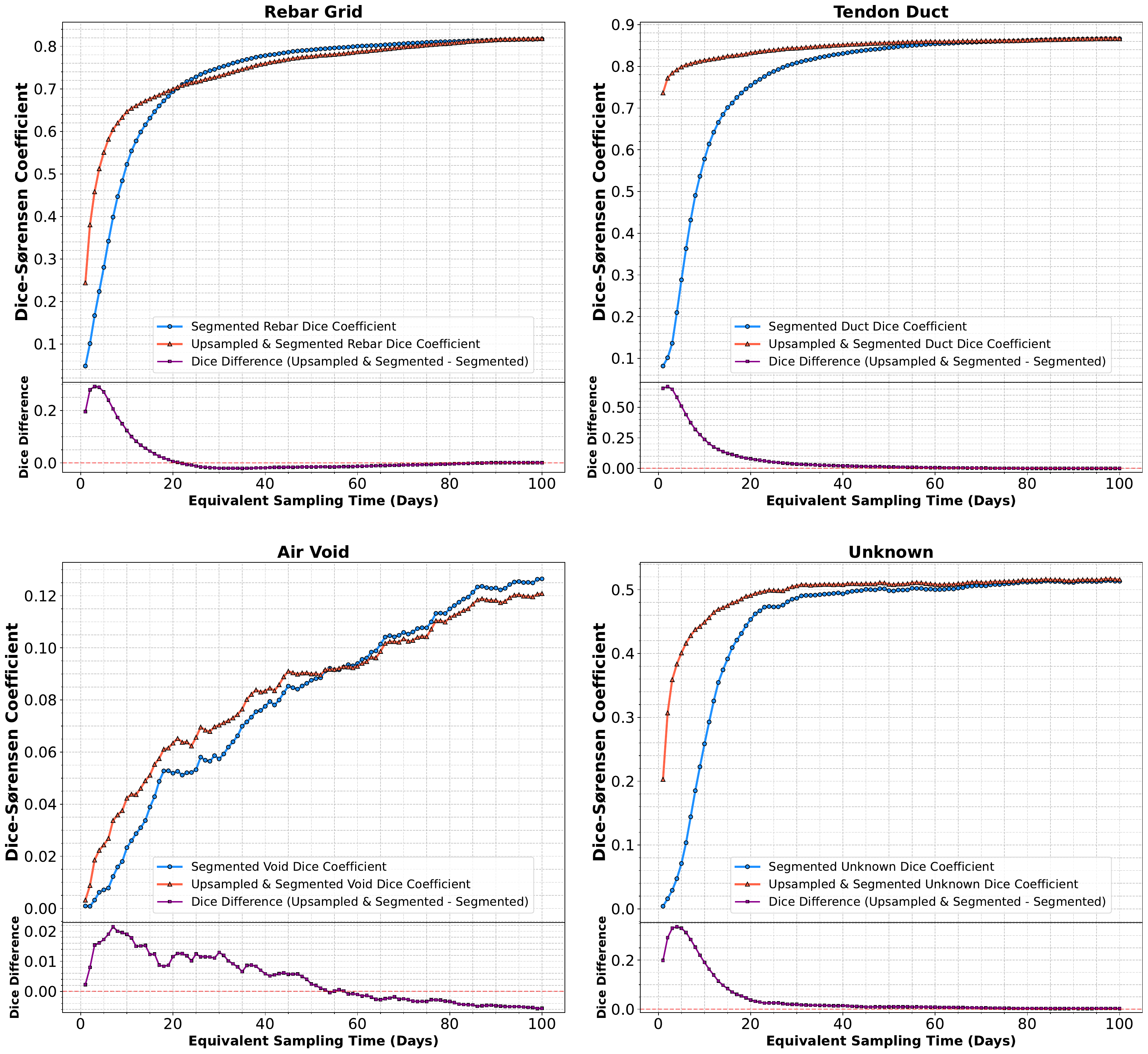} 
	\caption{Average Dice coefficient of the test dataset, calculated at each equivalent sampling time, for~each of the four segmentation object labels (rebar grid, tendon duct, void, unknown). Performance of the original input dataset is shown by blue circular points, where upsampled outputs are displayed with red triangular points. The~relative improvement of the upsampling model for a particular label is indicated by Dice difference below each object plot, shown in~purple.\label{dice_eval}}
\end{figure}

Using standard evaluation metrics to compare the quality of two images gives a limited interpretation of how the model improves the perception of the features within the muographic images.
Additionally, insufficient feature distinction results in the background pixels (attributed to concrete) in the images distorting the evaluation metrics by overwhelming the signal from other classes.
In order to evaluate the effect of the upsampling model on each feature type, we trained a second model to perform semantic segmentation of these images, where each pixel in the image is classified as one of the five predefined object labels: concrete, rebar grid, tendon duct, air void, or unknown.
If we were using experimental data, we would have to manually label images to retrieve ground truths; however, as~we were using Geant4 simulations, the~`true' position and composition of each object is known, so we used this to generate our ground truth segmentation images.
The segmentation model serves as an independent evaluation metric to assess the performance of the upsampling model. 
To minimise potential bias and maintain objectivity, this model was trained exclusively on high sampling time (100-day) muography images, deliberately excluding upsampler outputs from the training process.
The model performance was quantified using the Dice coefficient, which measures the overlap between pixelwise label predictions
 and 
 ground truth values.
The five-class segmentation labels were one-hot encoded into discrete binary masks, enabling class-specific overlap calculations.
By applying the segmentation model to both original and upsampled muography images, we could assess the upsampler's effect on each structural class.
A comparative analysis of Dice coefficients between original and upsampled images, focusing on the four non-background structural classes, is presented in Figure~\ref{dice_eval}.

The accuracy of the segmentation model for each class is indicated by the performance of the 100-day dataset inputs, showing Dice coefficients for rebar grids, tendon ducts, air voids, and `unknowns' of 0.8173, 0.8663, 0.1265, and 0.5132 respectively.
For the Dice coefficient difference subplots (in purple), we observe that the use of the upsampling model resulted in an excellent improvement for the identification of tendon ducts, a~good improvement for rebar and unknowns, and~little improvement for air voids.
We observe
\vspace{-3pt}
\noindent trends similar to that of the SSIM metric in Figure~\ref{ssim_psnr}, where the upsampling model was most effective at lower sampling times before converging with scores similar to the inputs at around 50--70 days of~sampling.

\begin{figure}[H]
	\includegraphics[width=0.86\textwidth]{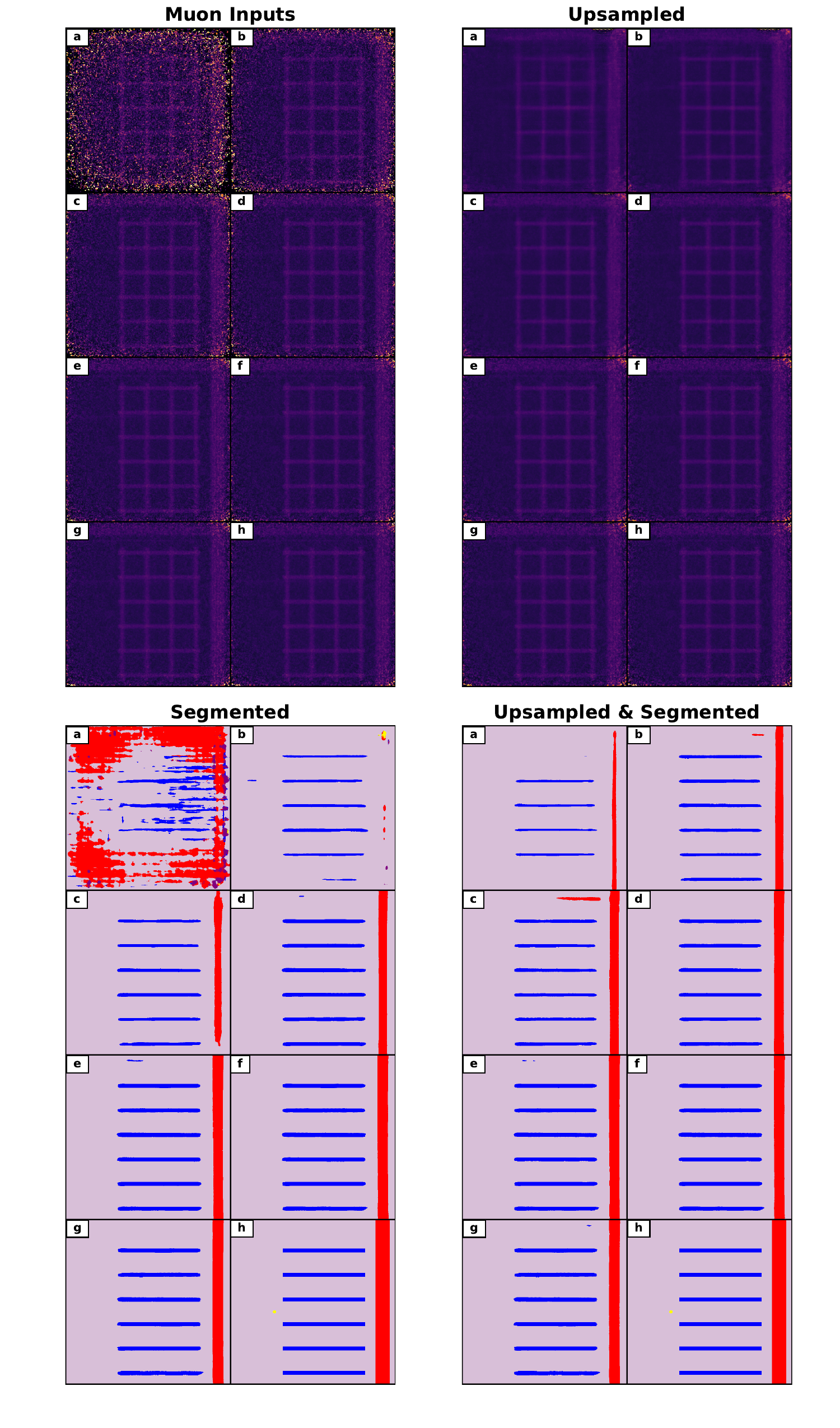}
	\caption{A single X--Y plane image slice for different equivalent sampling times: (\textbf{a}) 
 one day, (\textbf{b}) five days, (\textbf{c}) 10 days, (\textbf{d}) 20 days, (\textbf{e}) 40 days, (\textbf{f}) 60 days, (\textbf{g}) 80 days, (\textbf{h}) ground truth (100-day image for top panels, geometry truth for bottom panels). 
    These eight image versions are displayed as raw input (\textbf{top left}), upsampled (\textbf{top right}), segmented (\textbf{bottom left}), and~upsampled and segmented (\textbf{bottom right}).  
    Lilac, blue, red, and yellow indicate concrete, rebar, tendon ducts, and air voids, respectively.
    \label{XY_ims}}
\end{figure}

We can also perform visual inspection to obtain a better picture of how the segmentation model is performing, here using an image slice which is representative of model performance of on the test dataset.
Each of the four panels in Figure~\ref{XY_ims} show eight versions of a single X--Y plane, seven at different equivalent sampling times (a--g), and one as the ground truth (h).
The four panels show the original image, outputs of the upsampling model, outputs of the segmentation model, and outputs from a concatenation of both models.
The upsampled segmentation results (lower right) show a significant improvement at low sampling times in panels (a) and (b), aligning with trends shown by the Dice coefficients in Figure~\ref{dice_eval}.
A key feature of the segmentation model is the lack of y-aligned rebar features present. 

Looking at the segmentation ground truths in panel (h), we see that the segmentation model was correct in ignoring the y-aligned values, as indicated by the muography images.
This shows that the segmentation model has identified the y-aligned values as a shadowing effect, hence overcoming the z-smearing limitation of muography and ultimately improving the reliability of muography as a technique for accurate concrete~analysis.

\section{Discussion}
The results presented in this study demonstrate promising advancements in applying machine learning techniques to muography image enhancement and analysis.
Our findings show that cWGAN-GP models can effectively reduce required sampling times while maintaining image quality, with~particularly strong performance in feature detection for larger structural elements.
The following subsections examine the model performance in detail, addressing both its capabilities and limitations for both upsampling and segmentation~tasks.

\subsection{Image Upsampling~Capabilities}
The upsampling model effectively improved image quality, as demonstrated by both the SSIM and PSNR metrics in Figure~\ref{ssim_psnr}.
Looking at the input images (blue circular points), both metrics showed rapid improvement for equivalent sampling times between one and 35 days.
The SSIM followed a logarithmic trend that aligns with the feature detection performance shown in Figure~\ref{dice_eval} for rebar, tendon ducts, and 'unknown' objects, reflecting its design to quantify perceptual similarity.
In contrast, the~PSNR, which is calculated as a logarithm of the mean squared error (MSE), shows a different behaviour: after the initial 30-day period, it increased linearly until approximately 75 days before tending to infinity.
This difference arises because the PSNR is more sensitive to pixel-level variations, while the SSIM captures perceptual changes.
The combined analysis suggests that 30--40 days of sampling is sufficient to resolve large-scale features and reduce noise, with~additional sampling time primarily improving pixel-level detail that may not significantly impact image~perception.

The upsampled outputs (red triangular points) exhibited systematically higher SSIM and PSNR values compared to the original inputs at lower sampling times while following similar overall trends.
However, this advantage diminished as the sampling time increased, with~upsampled and original images showing identical SSIM performance at around \linebreak  50 days---the same point at which the corresponding input data (blue circular points) began to show significant diminishing returns.
This convergence can be understood from the data processing used: each image pixel value corresponds to a specific voxel in its z-plane layer, with~longer sampling times allowing more muon hits per voxel.
As the sampling time increased, these voxel values gradually approached their `true' values, and~the 50-day convergence point suggests sufficient sampling had been achieved to closely approximate these true values.
The PSNR of the upsampled images converged with its corresponding input images much later, at~around 80--85 days, before~the input metrics began to outperform.
This was likely caused by the lack of reconstruction of exact pixel-by-pixel noise variations present in the 100-day images, for~example, as shown in Figure~\ref{upsampling_ims}, where the edges and corners exhibited some noise fluctuation that was correspondingly more smooth in the upsampled 1-day~image.

Visual inspection of the images in Figure~\ref{upsampling_ims} and the top panels from Figure~\ref{XY_ims} can confirm the results exhibited from dataset-wide analysis.
At low sampling times, the~noise in the input images was significant enough to drown out most human-perceivable detail, combined with low or non-existent counting statistics around image edges caused by detector acceptance.
However, their upsampled counterparts exhibited drastic feature enhancement, combined with a significant reduction in noise to output a smooth but detailed image.
While the upsampling of 1-day images demonstrated remarkable improvement, some limitations remain evident---particularly in regions of very low statistics where noise fluctuations dominated.
For example, gaps appearing in rebar grids remained unfilled, despite our dataset containing only intact grids, due to the limited context window (4 $\times$ 4 pixels) of the convolutional~approach.

\subsection{Feature Segmentation~Performance}
Building on the image quality metrics discussed above, segmentation analysis provides deeper insight into how the upsampling model affects different structural features within the concrete samples.
By examining the Dice scores for each object type, we can quantitatively assess how well geometric features are preserved and enhanced through the upsampling~process.

Figure~\ref{dice_eval} shows a breakdown of the average segmentation performance for each object feature as a function of equivalent sampling time.
The rebar segmentation scores performed well, but early convergence occurred between the original and upsampled images at around 20 days---notably earlier than the convergence points for the SSIM, PSNR, tendon ducts, and `unknowns'.
Also noted is that between 20 and 85 days, the~original images slightly outperformed the upsampled ones, suggesting the upsampling model was negatively degrading rebar features.
Early convergence and outputs being outperformed by inputs suggest the model was washing out some rebar features.
This degradation likely stems from the varying dimensions of rebar in our dataset: grids range from 8 to 25 mm, with~the thinnest rebar being only four pixels wide at 2 mm pixelization.
The model may be struggling to preserve thin rebar features or mistaking them for shadows of larger ones.
The cylindrical grid structure of the rebar compounds this challenge, as~segmentation in the X--Y plane alone makes edge detection particularly difficult.
Visual inspections of samples containing thin rebar (8--10~mm) revealed poor segmentation performance; however, a comprehensive breakdown would confirm if this is the true source of the~issue.

The tendon ducts performed very well for the input data; additionally, 
the integration time to achieve the same results with the upsampling model was much improved.
This high performance is attributed to the large diameter and pixel volume of the objects.
Visual inspection of the dataset also revealed that the tendon ducts tended to be placed nearer the edges in low-statistics regions also attributing to a high relative performance gain---as the large size of the objects means that they are easily detectable through the high noise.
The placement of these ducts near the edges is caused by the concrete sample placement logic (where they are always placed after any rebar grids), which will on average tend to occupy the centres of blocks.
Bundling all three tendon duct features together will have also resulted in better object~identification.

The air voids were very poorly segmented.
The primary causes are their challenging detection in scattering tomographs due to low scattering statistics and their small spherical nature, which constitutes a low pixelwise percentage of the total dataset.
Visual inspection of the muography images revealed that most voids greater than 50~mm in diameter can be perceived. For~larger voids at~least, the~low Dice scores seem to stem from class imbalance.
During training, the~Dice component of the loss function (Equation~(\ref{custom_loss})) was calculated as the mean of the four object components.
However, due to the class imbalance---most notably in the case of air voids---the model is not incentivised to improve the correct detection of these pixels.
Their low representation in the dataset means they contribute minimally to the loss function, preventing significant improvement in their segmentation.
This issue can be addressed by setting custom Dice class weightings such that they reflect the proportion of pixels that each class occupies in the test~dataset.

The performance of the `unknown' object class was generally satisfactory but leaves room for improvement.
This class is particularly challenging to analyse due to its significant variety and density variations.
High-density materials, such as uranium and lead, tend to exhibit greater smearing, which can negatively impact Dice scores by increasing false positives.
Conversely, low-density materials, such as air and water, also contribute to lower Dice scores due to their propensity for false positives.
The purpose of this class is to provide the model with a miscellaneous category for objects that do not fall into the predefined categories of rebar, tendon ducts, or~air voids.
Further analysis of this class, broken down by material composition and size, would provide a clearer understanding of how objects with varying densities perform.
Additionally, evaluating models using objects with unseen densities and shapes may provide better insight to the success of this class'~objective.

Visual representations of the X--Y plane segmentation outputs are shown in Figure~\ref{XY_ims}, revealing several challenges.
The noisy original 1-day image struggled to perform any form of accurate segmentation map---note that it is predicting tendon duct pixels around the edges; the model is `guessing' tendons to exist in low-sample edge regions.
Low-sampled upsampled images (panels (a) on the right hand side) which were absent of high noise fluctuations showed a much better performance; however, they seemed to struggle to segment all rows of rebar despite being clearly visible.
This lack of spatial context limits the model's ability to accurately interpret the full geometry of the objects (in complex structures like cylindrical tendon ducts and rebar grids), where straight-edge detection and thicknesses are slightly lacking.
In addition to these issues, there are a small number of artefacts exhibited at the top of the images where the shadow of an x-aligned tendon duct is present.
Many of these are caused by the limited context size of the 4~$\times$~4 convolutional kernels, which are unable to make predictions based on pixels outwith this~area.

\subsubsection*{A Solution to Smearing~Artefacts}
We define our inverse imaging problem in muography as the imprecision of a scattering point in a direction orthogonal to the detector planes caused by the lack of information to place said scattering point on a given muon's line path.
For a detector plane parallel to the ground, the~muon path between the two sets of detectors will be along the z direction, with~the increased uncertainty in the z axis resulting in objects appearing to `smear' along this direction---also referred to as `shadows' or `shadowing'.
We can see in Figure~\ref{XY_ims} that the segmented images only contain x-aligned rebar, despite the associated muon images appearing to contain a rebar grid of y- and x-aligned rebar.
This is due to the segmentation model being trained to reproduce geometric ground truths that were absent of any smearing artefacts, learning the ability to accurately differentiate between shadows and `true' objects despite a lack of information along the~z axis.

We can investigate this phenomenon further by looking at a concrete block sample side-on by~stacking all segmented X--Y plane images back into a three-dimensional volume and taking a slice along the X--Z plane.
An example of a side-on slice showing the original, upsampled, segmented, and~upsampled and segmented versions at different equivalent sampling times is shown in Figure~\ref{XZ_ims}.
Despite the model not having any context in the z plane, segmentation outputs (bottom two panels) show remarkable improvements in object clarity when compared to the muographic images.
The upsampling model also exhibits a clear improvement to the low sampling times in panels (a) and (h).
However, this ability to reduce smearing effects to accurately distinguish smearing artefacts from actual object geometry would be much improved by using a model input which encompasses all three~dimensions.

For simplicity and computational efficiency, our current model processes each X--Y plane layer independently.
While effective, this method's primary limitation is the model's lack of contextual awareness of a layer's position within the overall scan, and it cannot utilise information from adjacent layers.
However, several methods exist to incorporate 3D context without processing full volumes.
These include orthogonal plane processing (where separate models are trained on X--Y, X--Z, and~Y--Z planes), multi-layer processing (which analyses small stacks of adjacent layers simultaneously) and patch-based processing (which splits the volume into manageable 3D sub-volumes).
These approaches would provide three-dimensional context while maintaining reasonable computational requirements and model~complexity.

\subsection{Future Directions: Model Refinement and Defect Detection~Applications}
The dataset used to train both models covered a wide range of sampling times.
The upsampling performance on the lowest generated image data was much better than anticipated, whereas performance at equivalent sampling times greater than 50--60 days yielded diminishing returns as well as indications of convergence between the input and output metric scores. 
Reducing the maximum sampling time would drastically decrease the required computation time for simulations and PoCA algorithm processing.
The high performance of the upsampling model on a 1-day equivalent sampling time, especially shown by the visual examples, indicates that the limits of the model should be further pushed to provide results for equivalent sampling times on the order of~hours.

Despite using segmentation purely as an analysis tool, its ability to resolve true object boundaries and therefore decrease smearing effects is a significant result.
Furthermore, it was noted that the segmentation performance on the upsampled images was limited, despite there being clear objects present.
This may be a symptom of using a two-model approach, as~the outputs of the upsampling model $\hat{y}$ are similar but not identical to the input $y$ used to train the segmentation model.
In order to achieve the most accurate outputs, without~the complexity of two models, we can simplify the task to a one-model approach which uses the true object geometries to generate a ground truth---whether that be a segmentation map of labels or~of material-specific values such as density or radiation~length.

The limited receptive field of convolutional kernels restricted the model to having a localised receptive field---prevalent in the non-uniformity of the rebar grids or straight edges of the segmented images.
While this issue can be circumvented through the use of hierarchical dilated convolutions~\cite{chen_encoder-decoder_2018}, allowing for a slight increase in context size, the~most popular solution is to use attention mechanisms~\cite{vaswani_attention_2017}.
Attention is a more recent and increasingly prominent methodology for model construction, powering state-of-the-art models in both language and vision tasks. It uses a tokenization system that splits images into patches, which are flattened and processed as individual tokens.
These tokens are positionally embedded to preserve spatial relationships, enabling the model to account for location dependence.
Unlike convolutions, which have a limited receptive field that grows with depth, attention mechanisms inherently consider relationships between all tokens, providing a global context at every layer.
This ability to capture long-range dependencies and spatial relationships makes attention a powerful alternative for vision~tasks.

While the concrete samples used to train the models are of realistic design, the~dataset poorly reflects a wide variety real-world scenarios which include more varied object orientations, concrete thicknesses, and detector orientations. 
Therefore, future dataset generation should aim to vary these parameters to allow the model to generalise to different~scenarios.

This study has looked at the upsampling and identification of objects commonly found within concrete interiors.
However, the~identification of objects solves only part of the problem pertaining to the characterization of built infrastructure.
The next step is to alter the model's objective to look toward not only identifying object location and material but~to also accurately identify defects---such as tendon duct strand corrosion, honeycombing, and air~voids.

Despite the general agreement of Geant4 simulations and real data for muography, it is important to verify these models on real-world data.
Therefore, testing of the models on data gathered from Niederleithinger~et~al.~\cite{niederleithinger_muon_2021}, as~well as planned future muography scans by Lynkeos Technology Ltd, will be important for validation of the results of this~study.
\begin{figure}[H]
	\includegraphics[width=0.8\dimexpr1.2\textwidth\relax]{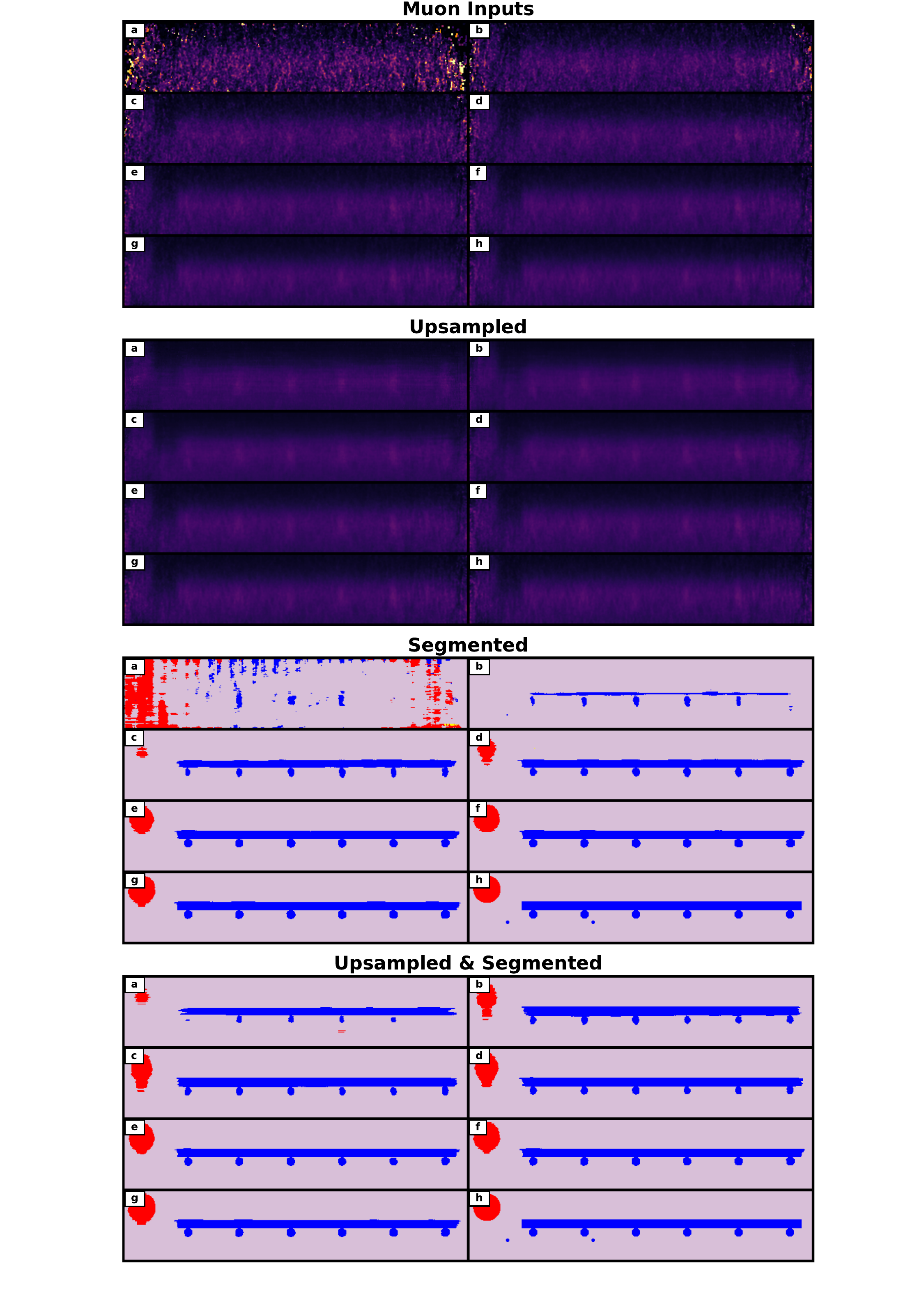}
	\caption{A 
 single X--Z plane 
 vertical image slice for different equivalent sampling times: (\textbf{a}) one day, (\textbf{b}) five days, (\textbf{c}) 10 days, (\textbf{d}) 20 days, (\textbf{e}) 40 days, (\textbf{f}) 60 days, (\textbf{g}) 80 days, (\textbf{h}) ground truth (100-day image for the top two panels, geometry truth for the bottom two panels). 
    These eight image versions}

    \label{XZ_ims}
\end{figure}

\begin{figure}[H]\ContinuedFloat

    \caption*{ are displayed as raw input (\textbf{first}), upsampled (\textbf{second} down), segmented (\textbf{third} down), and~upsampled and segmented (\textbf{bottom}).
    Lilac, blue, and red indicate concrete, rebar, and tendon \mbox{ducts, respectively}.}

\end{figure}
\unskip

\section{Conclusions}
In this study, we have demonstrated the successful application of deep learning techniques to address two significant challenges in muon scattering tomography: reducing required sampling times and improving feature identification in concrete structures. 
Our cWGAN-GP model achieved remarkable upsampling performance, with~1-day sampled images being enhanced to match the quality metrics of 21 baseline images for the SSIM and 31 baseline images for the PSNR.
This represents a significant reduction in the required sampling times for achieving high-quality muography images.
To evaluate the model's performance beyond traditional metrics, we developed an approach using semantic segmentation as an assessment tool.
This method revealed that the upsampling model's effectiveness varies significantly across different structural features, with~particularly strong performance in enhancing the visibility of tendon ducts and moderate improvements in rebar grid detection. 
The segmentation model also demonstrated an unexpected capability to differentiate between real features and shadowing artefacts caused by a low vertical resolution, thus addressing the inverse imaging problem inherent to muography.
While our results are promising, key challenges remain.
The model's performance on air void detection requires improvement, likely through the implementation of class-weighted loss functions to address class imbalance.
Additionally, the~current approach of processing 2D slices independently limits the model's ability to utilise full 3D contextual information, suggesting potential improvements through multi-plane or patch-based processing methods.
The transition from simulation to real-world application represents the next critical step in this research.
Future work will focus on generalizing the models to a range of realistic imaging scenarios, as~well as validating these models with experimental data and extending the segmentation capabilities to identify specific structural defects, such as tendon duct corrosion and concrete honeycombing.
These advancements could significantly impact the practical application of muography in infrastructure inspection, potentially reducing inspection times while improving the accuracy of defect~detection.

\vspace{6pt} 

\authorcontributions{Conceptualization, W.O., D.M., G.Y., and S.G.; methodology, W.O.; software, W.O. and G.Y.; validation, W.O.; formal analysis, W.O.; investigation, W.O.; resources, W.O.; data curation, W.O.; writing----original draft preparation, W.O.; writing----review and editing, D.M. and G.Y.; visualization, W.O.; supervision, D.M., G.Y., and S.G.; project administration, D.M.; funding acquisition, D.M. All authors have read and agreed to the published version of the~manuscript.}

\funding{This 
 research was funded through an agreement between Lynkeos Technology Ltd. and the College of Science and Engineering at the University of~Glasgow.}

\institutionalreview{Not applicable.}

\dataavailability{The data are available upon request.}

\acknowledgments{We give additional 
 thanks to Ernst Niederleithinger at BAM for his support and technical expertise, which informed the simulation~design.}

\conflictsofinterest{The 
 authors declare no conflicts of~interest.} 

\begin{adjustwidth}{-\extralength}{0cm}
\reftitle{References}


\PublishersNote{}
\end{adjustwidth}
\end{document}